# Echoes of Gravity

Douglas Scott & Martin White

*Center for Particle Astrophysics
and Department of Astronomy,
University of California, Berkeley, CA 94720-7304*

The study of anisotropies in the Cosmic Microwave Background radiation is progressing at a phenomenal rate, both experimentally and theoretically. These anisotropies can teach us an enormous amount about the way that fluctuations were generated and the way they subsequently evolved into the clustered galaxies which are observed today. In particular, on sub-degree scales the rich structure in the anisotropy spectrum is the consequence of gravity-driven acoustic oscillations occurring before the matter in the universe became neutral. The frozen-in phases of these sound waves imprint a dependence on many cosmological parameters, that we may be on the verge of extracting.

The measurement of Cosmic Microwave Background (CMB) temperature anisotropies by the *COBE* satellite at angular scales $\gtrsim 7°$ caused great excitement when announced in 1992 [1]. Subsequently there have been no fewer than 15 claimed detections in different regions of the sky by some 9 independent experiments, 4 balloon- and 5 ground-based [2]. While none of these data match the quality of the *COBE* results, they are at smaller angular scales, which means they probe different physical processes.

In the standard Hot Big Bang paradigm, the CMB photons that we observe suffered their last interaction with matter (Compton scattering off free electrons) at a redshift $z \simeq 1000$. The horizon at that time subtends an angle of $\sim 1°$ on the sky today. Any causal physical process operating at $z = 1000$ can have had no significant effect on angular scales $\gtrsim 1°$, but on scales less than $1°$, we are sensitive to a rich class of physical processes. In particular, gravity has had a chance to operate on these scales, imprinting a characteristic signature that depends on (amongst other things) the total matter content of the universe $\Omega_0$, the baryonic content $\Omega_B$, and the Hubble constant $H_0$. This signature can be seen in the features of the angular power spectrum of temperature anisotropies at sub-degree scales ($\ell \gtrsim 100$, see Figure 1). The detailed shapes of these features will be measurable in the future, and, since each multipole $\ell$ is an independent datum, there are potentially hundreds or even thousands of constraints available.

The physics that leads to these peaks has previously been obscured by their conventional name. In analogy with the features seen in the CMB spectrum for a universe that never became neutral, the structures seen in Figure 1 have been called 'Doppler peaks'. This conjures up an image involving Doppler shifts of photons scattering off moving electrons, which is the source of a qualitatively similar 'peak' of lesser amplitude in the fully ionized universe. The true explanation of the Doppler peaks for a standard thermal history is much more enlightening, and explains why there is so much to be learned from the detailed structure of the spectrum.*

---

* The underlying physics of these peaks, can be traced back through early analytic [3,4] and numerical [5,6] studies. The first accurate calculation, in the context of modern cosmological models, was presented in a pioneering paper [7] by Bond & Efstathiou. Recently, with the influx of data to spur interest, several authors have improved on the analytic approximations [8,9]. See also Refs. 10 and 11.

The sphere around us at $z \simeq 1000$, which represents the position at which the photons seen today last interacted directly with the matter, is called the last scattering surface. On large (angular) scales, one sees the temperature perturbations caused by photons climbing out of potential wells on this surface [12]. A long wavelength density or potential perturbation gives rise to a long wavelength temperature perturbation across this sphere and hence a large angular scale temperature anisotropy. Additional anisotropies on these scales can be imprinted upon the photons as they travel toward us, by time dependent metric perturbations, which change the energy of the photons by an amount proportional to the integral of $\dot{g}_{\mu\nu}$ along the line of sight. However, the peaks in the power spectrum around $\ell \simeq 200$ ($\theta \simeq 30'$) are generated by physical processes occurring 'in' the surface of last scattering, rather than being reflections of fluctuations imprinted on the surface by the initial conditions.

To understand these features, consider the universe just before it cooled enough to allow protons to capture electrons. At these early times, the photons and proton-electron plasma (the baryons) were tightly coupled by Compton scattering and electromagnetic interactions. These components thus behaved as a single 'baryon-photon fluid' with the photons providing the pressure and the baryons providing inertia. In the presence of a gravitational potential, there is a driven acoustic oscillation in the baryon-photon fluid. As the universe ages and the horizon grows, modes of larger and larger wavelength 'come inside' the horizon. Consider an overdensity that is just coming inside the horizon at a given time. This mode begins to collapse under its own gravity, becoming more overdense, until the photon pressure support becomes sufficient to halt the collapse. At this point, the overdensity 'rebounds', setting up an oscillation. The oscillation is described by a driven harmonic oscillator equation, with the driving force given by gravity, the inertia by the baryons and the restoring force (pressure) by the photons.

When the universe becomes cool enough for the protons to capture the electrons and form neutral hydrogen, the photon-baryon cross section drops precipitously, allowing the baryonic matter and the radiation to go their separate ways. With the release of their pressure support, the baryons fall into the dark matter potential wells and their perturbations grow to form the structures we see today. The photons, meanwhile, propagate to the observer, interacting with the matter in the universe only through gravity. The energy density, or brightness, fluctuations in the photons (coming from density and velocity com-ponents of the harmonic oscillator) are seen by the observers as temperature anisotropies. A mode which was maximally overdense at the time when the universe recombined will appear brighter and hotter than average, while those oscillations which were maximally underdense will appear colder. Modes which are passing through a density-null have only kinetic energy and so their brightness will be reduced by an amount proportional to the sound speed squared.

Given this discussion, it is possible to understand the peaks in Figure 1. The first peak corresponds to a perturbation that crossed the horizon and became maximally overdense just when the universe recombined, i.e. $1/4$ of an oscillation. Similarly the third, fifth etc. peaks represent scales that have undergone an extra integral number of oscillations by this time. The even peaks are maximally underdense modes, which also give rise to power peaks (power is temperature difference *squared*), though of generally smaller amplitude since the rebound must fight against gravity. The troughs, which do not extend all the way to zero, are velocity maxima, which are $\pi/2$ out of phase with the density maxima.

Of course the recombination event is not instantaneous, and it is this finite duration which leads to the damping of the anisotropies on the smallest scales. In the standard scenario, the thickness of the last scattering surface is $\sim 100$ in redshift. Two damping effects operate here. The first is that in the time the universe takes to recombine, photons can random walk a certain distance. Perturbations on scales smaller than this distance thus lose their photon anisotropies by diffusion [13]. The second effect comes from anisotropies with wavelength smaller than the thickness destructively interfering across the last scattering surface, leading to power-law suppressed anisotropies [14]. One can see that this suppression, plus general damping of oscillations with the expansion, leads to successively smaller peaks as $\ell$ increases.

The beauty of this picture is that the precise shape of the anisotropy spectrum depends on details of the gravito-acoustic oscillations, which leads to observable signatures of cosmological parameter variations. The amplitude of the peaks clearly depends on the ratio of the restoring force to the inertia, or on the baryon to photon ratio. With the temperature of the CMB now determined very precisely [15], the photon energy density as a function of redshift is well known. The baryon number density depends on $\Omega_B h^2$, where $h$ is $H_0$ in units of $100\,\mathrm{km\,s^{-1}\,Mpc^{-1}}$. Hence the peak heights are primarily a function of $\Omega_B h^2$. Of course, photons coming from an overdense region still need to climb out of the potential well associated with such an overdensity, and in the process they lose energy. The size of the potential depends on how close to the epoch of matter-radiation equality the last scattering event was, since potentials inside the horizon decay in a radiation dominated era but not in a fully matter dominated one. This introduces an additional dependence on $h$, which allows the degeneracy between $\Omega_B$ and $h^{-2}$ to be broken.

The detailed shapes, heights and locations of these peaks and troughs are firm predictions of models like CDM, and can tell us a wealth of information about such parameters as $\Omega_0$, $h$, $\Omega_B$ etc. Without reciting a litany of effects as various cosmological parameters or assumptions are varied, one point deserves special mention. The physical scale associated with the first peak is the horizon size at last scattering. Since the recombination process depends primarily on temperature, it occurs reliably at $z \simeq 1000$. The angle subtended by this physical scale, at redshift 1000, clearly depends on the geometry of the universe. If the universe is 'flat', then photon geodesics are straight lines. However if the universe is 'open', then geodesics diverge and a given physical scale at high redshift subtends a smaller angular scale today, by a factor of roughly $\Omega_0^{1/2}$. This means that the angular position of the first peak is a sensitive probe of the spatial curvature of the universe [16,17].

The current experimental situation is indicated in Figure 2 (for more details see Ref. 2). (The plotted quantity is a measure of temperature difference or the square root of 'power'.) If the power spectrum was in fact featureless, then the points would scatter about a horizontal line. While it is clearly early days for drawing firm conclusions based on degree-scale data, there is nevertheless the enticing impression of more power at $\ell \sim 200$ than at *COBE* scales. If we take these data at face value, we can make two statements: a model with a peak like that of standard Cold Dark Matter is a good fit to all the data; and a model with no peak is 'ruled out' at more than 99% confidence. But the main lesson is optimism about the possibilities for genuinely measuring cosmological parameters in the future.

At present, the experimental situation with regards CMB anisotropies is progressing at an extraordinary rate. The largest hurdles in the experimental work are control of systematics and obtaining sufficient observing time. While detector sensitivity is still an issue, it is becoming less crucial with time. Current observational programs continue to amass data and there are plans for long duration observations from the South Pole and balloon borne platforms. The next quantum leap in observations is likely to come from another satellite mission, however, where observing time and systematics are more favourable.

It is now clear that, in order to extract the most interesting science, such an experiment needs to have sensitivity at angular scales which get *over* the first peak. If a map can be made of a large fraction of the foreground-free sky, then we can extract all $C_\ell$'s from $\ell \sim 2$ up to roughly the inverse of the beam size. This information probes the processed and primordial power spectra together, which enables us to infer information about the generation *and* evolution of cosmological density perturbations. This means that we learn about both the physics of the very early universe ($t \lesssim 10^{-30}$s) and the physics at $z \sim 1000$ ($t \sim 300{,}000$ years).

We have made some preliminary investigations of the discriminatory power in a satellite mission that measures much of the sky at $0°\!.5$ resolution and with realistically achievable levels of noise (say $20\,\mu$K per pixel). We found that in many Monte Carlo realizations of an input model, we could recover the parameters, e.g. $\Omega_0$ and $\Omega_B$, to a relative accuracy of a few percent.

Other commonly discussed parameters will also lead to observable signatures: the Hubble constant, the cosmological constant, the primordial spectral slope, a background of long wavelength gravitational waves, etc. Although these may be difficult to separate precisely, we are anticipating information from several *hundreds* of multipoles. Hence we expect not only a measurement of these quantities, but also sensitivity to more detailed physics, e.g. the recombination process, the reionization of the universe, non-power-law initial conditions, or extra particle species.

All this will be possible because nature has provided us with an astonishing mine of information in the CMB anisotropies: the echoes of gravitational and pressure forces which operated at $z = 1000$. Already, it looks like there is a hint of this structure peeking out above the noise in the experimental data. If borne out, the presence of these gravito-acoustic peaks will represent a triumph for theoretical predictions of the Big Bang, through the physics of simple sound waves in the early Universe.

## ACKNOWLEDGEMENTS

We would like to thank Joe Silk, Naoshi Sugiyama, Wayne Hu and Ted Bunn for many useful conversations. This work was supported in part by grants from the NSF.

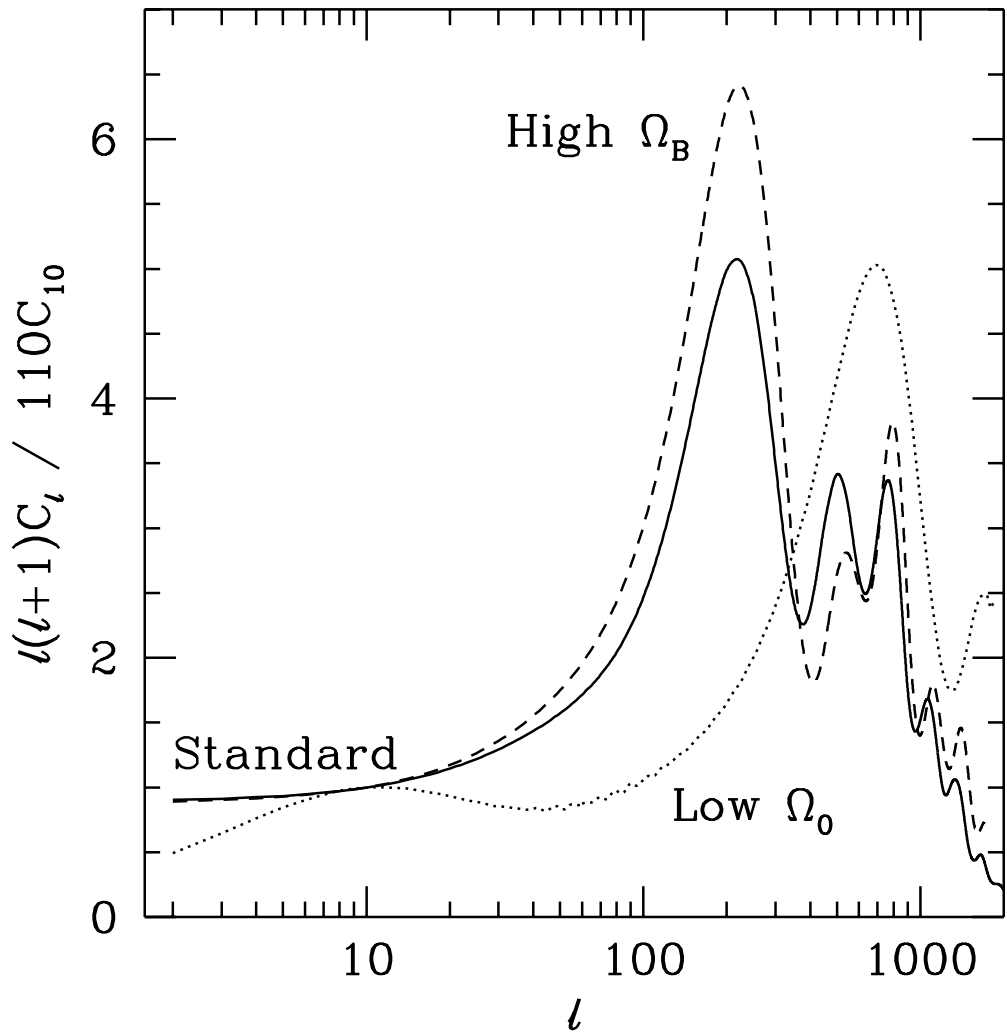

**Figure 1.** The spectrum of temperature fluctuations for a standard Cold Dark Matter model (solid: $\Omega_0 = 1$, $h = 0.5$ and $\Omega_B = 0.05$) and two variants. The dashed curve shows the effect of doubling $\Omega_B$, and the dotted curve is for $\Omega_0 = 0.1$. The quantity $\ell(\ell+1)C_\ell$ is power per logarithmic interval in multipole number $\ell \sim \theta^{-1}$. Note that the curve is fairly flat at small $\ell$ (large angular scales) and has considerable structure at larger $\ell$ (small angular scales), which comes from the frozen-in phases of gravity-driven acoustic oscillations. The relative heights and positions of the peaks give information about the cosmological parameters.

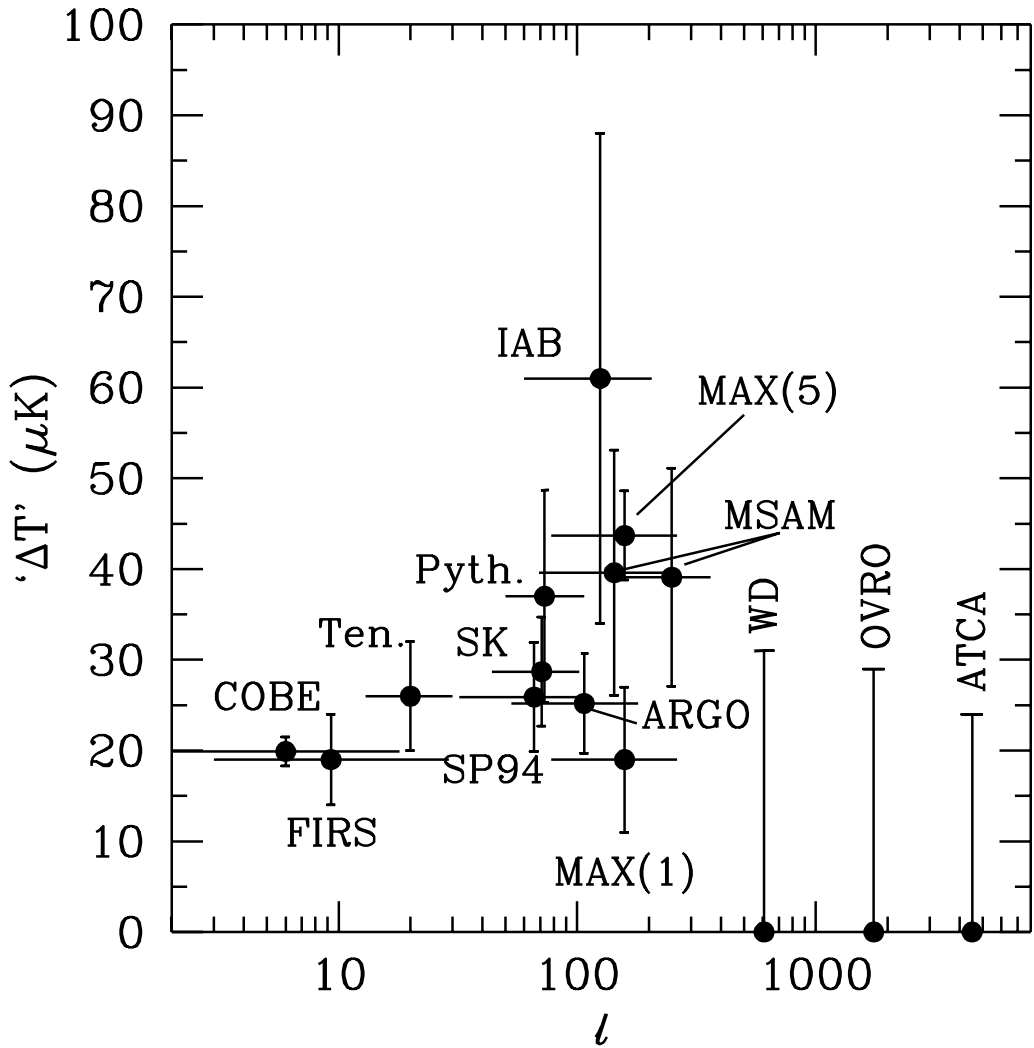

**Figure 2.** The amplitude of '$\Delta T$' fluctuations for published experimental data, as a function of scale (multipole $\ell \sim \theta^{-1}$). Each experimental result has been converted to a consistent measure of the rms anisotropy (see Ref. 11 for details). The vertical error bars are $\pm 1\sigma$, while the horizontal lines represent the range of scales to which the experiment is sensitive. There are also three smaller-scale results plotted as 95% upper limits. The general rise in the area around $\ell \simeq 200$ can be interpreted as evidence for a peak in the radiation power spectrum.